\documentclass[11pt]{iopart}
\usepackage{cite} 
\usepackage{wrapfig}
\usepackage{amsfonts}
\usepackage{amsbsy}
\usepackage{graphicx}
\usepackage{graphics}
\usepackage{hyperref}

\begin{document}

\title[The exact solution of a fragmented Bose-Hubbard model and twisted models]{The exact solution of a fragmented Bose-Hubbard model and twisted models}

\author{ Gilberto N. Santos Filho$^{1}$}

\address{$^{1}$ Universidade Federal de Sergipe - UFS \\
Departamento de F\'{\i}sica - DFI \\ 
Cidade Universit\'aria Prof. Jos\'e Alo\'{\i}sio de Campos \\
Av. Marechal Rondon, s/n, Jd. Rosa Elze, S\~ao Crist\'ov\~ao - SE - Brazil}
\ead{gfilho@cbpf.br, gnsfilho@gmail.com}


\begin{abstract}
We present the exact solution of a family of fragmented Bose-Hubbard models and represent the models as graphs with the condensates in the vertices. The models are solved by the algebraic Bethe ansatz method. We show that the models have the same spectrum of a family of twisted models that we get by a local $U(1)$ gauge transformation.
\end{abstract}

\section{Introduction}

  After the realization of Bose-Einstein condensates (BEC)\cite{bose,einstein}, achieved by taking dilute alkali gases to ultra low temperatures \cite{early,angly,wwcch,anderson,wk,Hulet,Bagnato,Yukalov} 
we have a fast growing of the investigations dedicated to the comprehension 
of  new phenomena  either in the experimental or theoretical domains. 
 The fast increasing of the control and production of Bose-Einstein condensates (BECs) in different geometries has permitted the study of these systems in different physical situations. The fragmentation of a BEC to produce a Josephson junction \cite{Josephson1,Josephson2} using BECs opened the possibility to study the quantum tunnelling of the atoms across a barrier between the condensates \cite{Ketterle1,Ketterle2,Oberthaler1,Oberthaler2,Oberthaler3}. There is also the possibility to produce optical lattices in one-dimension (1D), two-dimensions (2D) and three-dimensions (3D) using one, two or three orthogonal standing waves \cite{Bloch}.  Using superposition of light in different direction is possible to create any arbitrary trapping configuration as for example a ring or a superconducting quantum interference devices (SQUID) with an  atomic BEC \cite{Ryu1,Ryu2}. Another experimental realization of BECs fragmentation  is the two-legs bosonic ladder to study chiral current and Meissner effect \cite{ladder1,ladder2,goldman,goldman2,orignac,haller2,galitski}.  These experimental realizations have opened the possibility to introduce new models that allow to study the tunnelling of the atoms between the BECs in different configurations. Many of these models are exactly solvable  by the algebraic Bethe ansatz (ABA) method \cite{Roditi,jon1,jon2,jonjpa,key-3,dukelskyy,Ortiz,Kundu,eric5,GSantosaa,GSantos,GSantos11,multistates,Xin,Ymai,Bethe-states,rubeni} that open the possibility to take in account quantum fluctuations, allowing us go beyond the results obtained by mean field approximations to compare with simulation and experiments. It is important to note  that exactly solvable models has been applied to compare with experiments in the context of ultracold atomic-molecular physics \cite{batchelor2} and in nuclear magnetic resonance (NMR) \cite{screp,kino1,kino2,kitagawa,haller,liao,coldea,nmr1}. This method can furnish some new insights in 
this area, and contribute as well to  increase the field of integrable systems  \cite{hertier, batchelor1,batchelor3}. The algebraic formulation of the Bethe ansatz, and the associated quantum inverse scattering method (QISM), was developed in \cite{fst,ks,takhtajan,korepin,faddeev} by Faddeev and the Leningrad School\cite{faddeev2}. The QISM has been used to study different branch of physics, such as, condensed matter and statistical physics \cite{korepin1,yang,korepin2,ek,ek2},  high energy physics \cite{lipatov, korch, belitsky,Gromov1,Gromov2,CAhn,dorey}, conformal field theory \cite{blz},
 and quantum algebras (deformations of universal
enveloping algebras of Lie algebras) \cite{jimbo85,jimbo86,drinfeld,frt}.  

We are considering here the bosonic multi-states Lax operator  \cite{multistates} that permits to solve a family of models of fragmented BECs coupled by Josephson tunnelling. This Lax operator is a generalization of the bosonic Lax operator in \cite{Kuznet, key-3}, where a Lax operator is 
defined for a single canonical boson operator, but instead of a single operator we choose a linear combination of independent canonical boson operators.

Using a local $U(1)$ gauge transformation we construct twisted models of these fragmented BECs model and using an appropriated choice of the parameters we proof that these twisted models and the fragmented models have the same spectrum, changing only their wavefunctions. 

The paper is organized as follows. In section 2, we will review briefly the ABA method and present the Lax operators. In section 3, we present a family of models of fragmented BECs coupled by Josephson tunnelling. In section 4 we present a family of models of twisted fragmented BECs coupled by Josephson tunnelling.  In Section 5 we present the exact solution of the family of models of fragmented BECs coupled by Josephson tunnelling. In section 6, we show that the twisted models have the same spectrum of the fragmented BECs models but with different wavefunctions. In section 7, we summarize the results.

\section{Algebraic Bethe ansatz method}

In this section we will shortly review the ABA method and present the transfer matrix used to get the solution of the 
models \cite{jonjpa,Roditi}. We begin with the $gl(2)$-invariant $R$-matrix, depending on the spectral parameter $u$,

\begin{equation}
R(u)= \left( \begin{array}{cccc}
1 & 0 & 0 & 0\\
0 & b(u) & c(u) & 0\\
0 & c(u) & b(u) & 0\\
0 & 0 & 0 & 1\end{array}\right),\end{equation}

\noindent with $b(u)=u/(u+\eta)$, $c(u)=\eta/(u+\eta)$ and $b(u) + c(u) = 1$. Above,
$\eta$ is an arbitrary parameter, to be chosen later.

It is easy to check that $R(u)$ satisfies the Yang-Baxter equation

\begin{equation}
R_{12}(u-v)R_{13}(u)R_{23}(v)=R_{23}(v)R_{13}(u)R_{12}(u-v),
\end{equation}

\noindent where $R_{jk}(u)$ denotes the matrix acting non-trivially
on the $j$-th and the $k$-th spaces and as the identity on the remaining
space.

Next we define the monodromy matrix  $\hat{T}(u)$,

\begin{equation}
\hat{T}(u)= \left( \begin{array}{cc}
 \hat{A}(u) & \hat{B}(u)\\
 \hat{C}(u) & \hat{D}(u)\end{array}\right),\label{monod}
\end{equation}

\noindent such that the Yang-Baxter algebra is satisfied

\begin{equation}
R_{12}(u-v)\hat{T}_{1}(u)\hat{T}_{2}(v) = \hat{T}_{2}(v)\hat{T}_{1}(u)R_{12}(u-v).\label{RTT}
\end{equation}

\noindent In what follows we will choose a realization for the monodromy matrix $\pi(\hat{T}(u))=\hat{L}(u)$  
to obtain solutions of a family of fragmented Bose-Einstein condensates models. In this construction, the Lax operators $\hat{L}(u)$  have to satisfy the relation

\begin{equation}
R_{12}(u-v)\hat{L}_{1}(u)\hat{L}_{2}(v)=\hat{L}_{2}(v)\hat{L}_{1}(u)R_{12}(u-v).
\label{RLL}
\end{equation}

Then, defining the transfer matrix, as usual, through

\begin{equation}
\hat{t}(u)= tr \;\pi(\hat{T}(u)) = \pi(\hat{A}(u) + \hat{D}(u)),
\label{trTu}
\end{equation}
\noindent it follows from (\ref{RTT}) that the transfer matrix commutes for
different values of the spectral parameter; i. e.,

\begin{equation}
[\hat{t}(u),\hat{t}(v)]=0, \;\;\;\;\;\;\; \forall \;u,\;v.
\end{equation}
\noindent Consequently, the models derived from this transfer matrix will be integrable. Another consequence is that the 
coefficients $\hat{\mathcal{C}}_k$ in the transfer matrix $\hat{t}(u)$,

\begin{equation}
\hat{t}(u) = \sum_{k} \hat{\mathcal{C}}_k u^k,
\end{equation}
\noindent are conserved quantities or simply $c$-numbers, with

\begin{equation}
[\hat{\mathcal{C}}_j,\hat{\mathcal{C}}_k] = 0, \;\;\;\;\;\;\; \forall \;j,\;k.
\end{equation}

If the transfer matrix $\hat{t}(u)$ is a polynomial function in $u$, with $k \geq 0$, it is easy to see that,

\begin{equation}
\hat{\mathcal{C}}_0 = \hat{t}(0) \;\;\; \mbox{and} \;\;\; \hat{\mathcal{C}}_k = \frac{1}{k!}\left.\frac{d^k\hat{t}(u)}{du^k}\right|_{u=0}. 
\label{C14b}
\end{equation}

 For the standard bosonic operators satisfying the Heisenberg-Weyl algebra 

\begin{equation}
 [\hat{p}_{i}^{\dagger},\hat{q}_{j}^{\dagger}] = [\hat{p}_{i},\hat{q}_{j}] = 0, \qquad [\hat{p}_{i},\hat{q}_{j}^{\dagger}] = \delta_{pq}\delta_{ij}\hat{I},
\end{equation}
\begin{equation}
[\hat{N}_{pi},\hat{q}_{j}^{\dagger}]= +\hat{p}_{j}^{\dagger}\delta_{pq}\delta_{ij}, \qquad [\hat{N}_{pi},\hat{q}_{j}]= -\hat{p}_{j}\delta_{pq}\delta_{ij},
\end{equation}
\noindent with $p,q = a \;\mbox{or}\; b$, $i = 1,\ldots, n$ and $j = 1, \ldots, m$, we have the following Lax operators,

\begin{equation}
\hat{L}^{\Sigma_{a}^n}(u) = \left(\begin{array}{cc}
u\hat{I} + \eta\sum_{j=1}^{n}\hat{N}_{aj} & \sum_{j=1}^{n}t_{aj}\hat{a}_{j}\\
\sum_{j=1}^{n}s_{aj}\hat{a}_{j}^{\dagger} & \eta^{-1}\zeta_a \hat{I}
\end{array}\right),
\label{L2}
\end{equation}
\noindent and
\begin{equation}
\hat{L}^{\Sigma_{b}^m}(u) = \left(\begin{array}{cc}
u\hat{I} + \eta\sum_{k=1}^{m}\hat{N}_{bk} & \sum_{k=1}^{m}t_{bk}\hat{b}_{k}\\
\sum_{k=1}^{m}s_{bk}\hat{b}_{k}^{\dagger} & \eta^{-1}\zeta_b \hat{I}
\end{array}\right),
\label{L3}
\end{equation}
\noindent if the conditions, $\zeta_a = \sum_{j=1}^{n}s_{aj}t_{aj}$ and $\zeta_b = \sum_{k=1}^{m}s_{bk}t_{bk}$, are satisfied. The above Lax operators satisfy the equation (\ref{RLL}).

\section{The fragmented model}

The Hamiltonian of the BECs fragmented model is,

\begin{eqnarray}
\hat{H} & = & \sum_{j=1}^{n} U_{ajaj}\hat{N}^2_{aj} + \sum_{k=1}^{m} U_{bkbk}\hat{N}^2_{bk}  + \frac{1}{2}\sum_{ j=1 }^{n}\sum_{ l=1 (j\neq l)}^{n} U_{ajal} \hat{N}_{aj}\hat{N}_{al} \nonumber \\ 
&+& \frac{1}{2}\sum_{ k=1 }^{m}\sum_{ s=1 (k\neq s)}^{m} U_{bkbs} \hat{N}_{bk}\hat{N}_{bs}\nonumber \\ 
&+& \sum_{j=1}^{n}\sum_{k=1}^{m} U_{ajbk} \hat{N}_{aj}\hat{N}_{bk} + \sum_{j=1}^{n} (\epsilon_{aj}  - \mu_{aj})\hat{N}_{aj} + \sum_{k=1}^{m}(\epsilon_{bk} - \mu_{bk})\hat{N}_{bk} \nonumber \\ 
&-& \sum_{j=1}^{n}\sum_{k=1}^{m} J_{ajbk}(\hat{a}_{j}^{\dagger}\hat{b}_{k} + \hat{b}_{k}^{\dagger}\hat{a}_{j}).
\label{H1}
\end{eqnarray}
\noindent The parameters, $U_{prqs}$, describe the atom-atom $S$-wave scattering, the $\mu_{ps}$ parameters are the external potentials and  $\epsilon_{ps}$ are the energies in the BECs. The parameters $J_{ajbk}$ are the tunnelling amplitudes. The operators $N_{ps}$ are the number of atoms operators. The labels $p$ and $q$ stand for the BECs $a$ and $b$ with $r,s=j,k$, $j = 1,\ldots,n$ and $k = 1,\ldots,m$. We just remark that $U_{pjpk} = U_{pkpj}$. The BECs are coupled by Josephson tunnelling and the total number of atoms, $\hat{N} = \sum_{j=1}^n \hat{N}_{aj}  + \sum_{k=1}^m \hat{N}_{bk}$, is a conserved quantity, $[\hat{H},\hat{N}] = 0$.

For each pair of $aj$ and $bk$ condensates we have the following currents algebra \cite{CA-Gil}

\begin{equation}
[\mathcal{T}_{ajbk},\mathcal{J}_{ajbk}] = i \mathcal{I}_{ajbk},\quad [\mathcal{I}_{ajbk},\mathcal{T}_{ajbk}] = i\mathcal{J}_{ajbk}, \quad [\mathcal{J}_{ajbk},\mathcal{I}_{ajbk}] = i\mathcal{T}_{ajbk},
\label{CA}
\end{equation}
\noindent with the currents

\begin{eqnarray}
\mathcal{I}_{ajbk} &=& \frac{1}{2}(N_{aj} - N_{bk}),  \label{Current-I}\\
\mathcal{J}_{ajbk} &=& \frac{1}{2i}(\hat{a}_{j}^{\dagger}\hat{b}_{k} - \hat{b}_{k}^{\dagger}\hat{a}_{j}),  \label{Current-J} \\
\mathcal{T}_{ajbk} &=& \frac{1}{2}(\hat{a}_{j}^{\dagger}\hat{b}_{k} + \hat{b}_{k}^{\dagger}\hat{a}_{j}). 
\label{Current-T} 
\end{eqnarray}

 The state space is spanned by the base $\{|n_{a1},\ldots, n_{bm}\rangle\}$ and we can write each vector state as 

\begin{eqnarray}
  |n_{a1},\ldots, n_{bm}\rangle &=& \frac{1}{\sqrt{\prod_{j=1}^n n_{aj}!\prod_{k=1}^m n_{bk}!}}\prod_{j=1}^n (\hat{a}_{j}^{\dagger})^{n_{aj}}\prod_{k=1}^m(\hat{b}_{k}^{\dagger})^{n_{bk}}|0\rangle, \nonumber\\
\label{state1}   
\end{eqnarray}

\noindent where $|0\rangle = |0_{a1},\ldots, 0_{bm}\rangle$ is the vacuum vector state in the Fock space. We can use the states (\ref{state1}) to write the matrix representation of the Hamiltonian (\ref{H1}). The dimension of the space of states increase very fast when we increase $N$,

\begin{equation}
  d = \frac{(n + m -1 + N)!}{(n + m -1)!N!},
\end{equation}
\noindent where $n + m$ is the total number of BECs in the system and $N$ is the eigenvalues of $\hat{N}$, $N = \sum_{j=1}^n n_{aj} + \sum_{k=1}^m n_{bk}$. In the case where we have only two BECs \cite{GSantosaa} (one $a_1$ and one $b_1$) the dimension is $d = N + 1$.

We can write a general state in the space spanned by the vectors (\ref{state1}) as

\begin{equation}
|\Psi\rangle = \sum_{j=1}^n\sum_{k=1}^m \frac{1}{\sqrt{\prod_{j=1}^n n_{aj}!\prod_{k=1}^m n_{bk}!}}\prod_{j=1}^n (\hat{a}_{j}^{\dagger})^{n_{aj}}\prod_{k=1}^m(\hat{b}_{k}^{\dagger})^{n_{bk}}|0\rangle.
\label{state1g}
\end{equation}
\noindent If $N$ is even, $\sum_{j=1}^n n_{aj}$ and $\sum_{k=1}^m n_{bk}$ have the same parity and the wavefunction (\ref{state1g}) is even. If $N$ is odd, $\sum_{j=1}^n n_{aj}$ and $\sum_{k=1}^m n_{bk}$ have different parity and the wavefunction (\ref{state1g}) is odd \cite{CA-Gil}. The kernel of the Hamiltonian (\ref{H1}) is the vacuum,

\begin{equation}
ker(\hat{H})={\bigotimes_{j=1}^{n}|0\rangle_{aj} \otimes \bigotimes_{k=1}^{m}|0\rangle_{bk}},
\end{equation}
\noindent and the dimension of the Hilbert space is $d + 1$.

\begin{figure}
\begin{center}
\begin{tabular}{cc}
$(a)$ & $(b)$ \\
\includegraphics[scale=0.35]{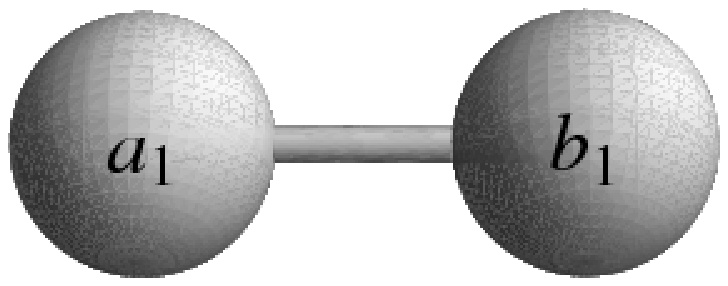} & \includegraphics[scale=0.35]{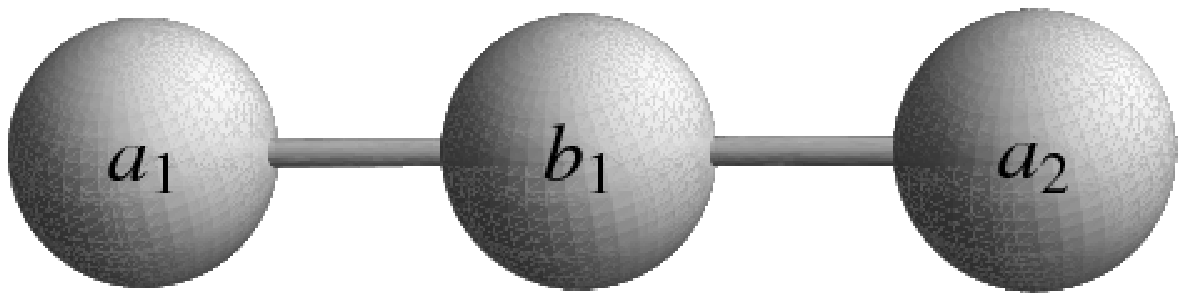} \\
$(c)$ & $(d)$ \\
\includegraphics[scale=0.4]{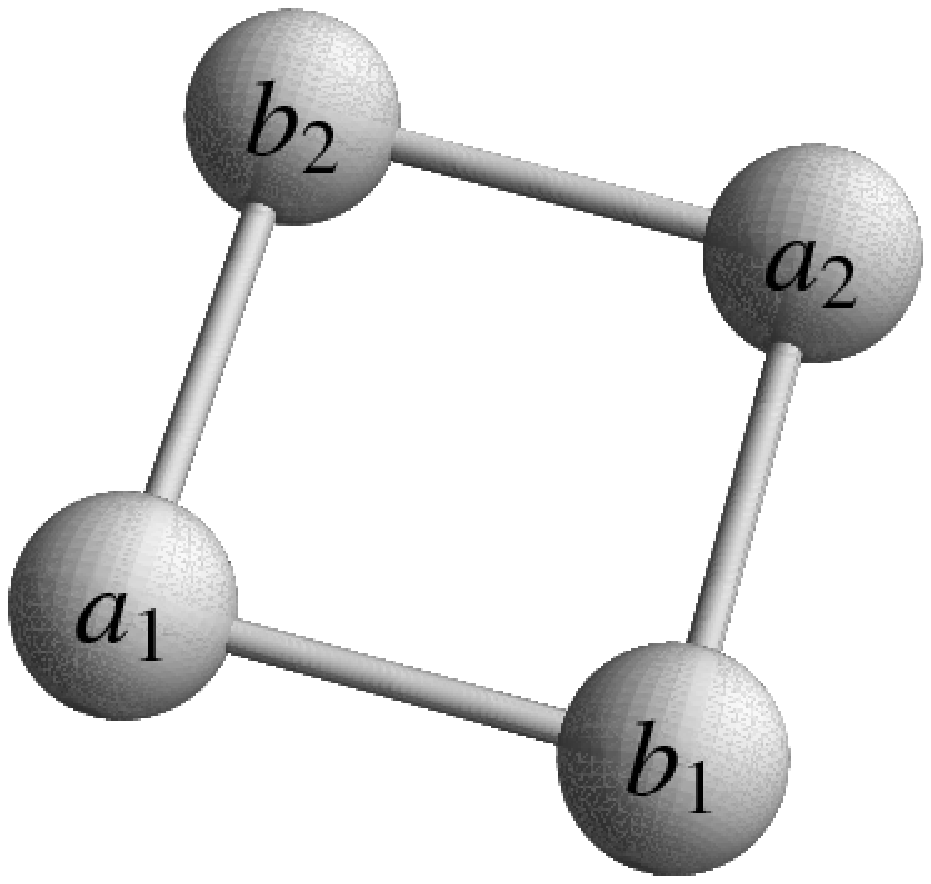} & \includegraphics[scale=0.4]{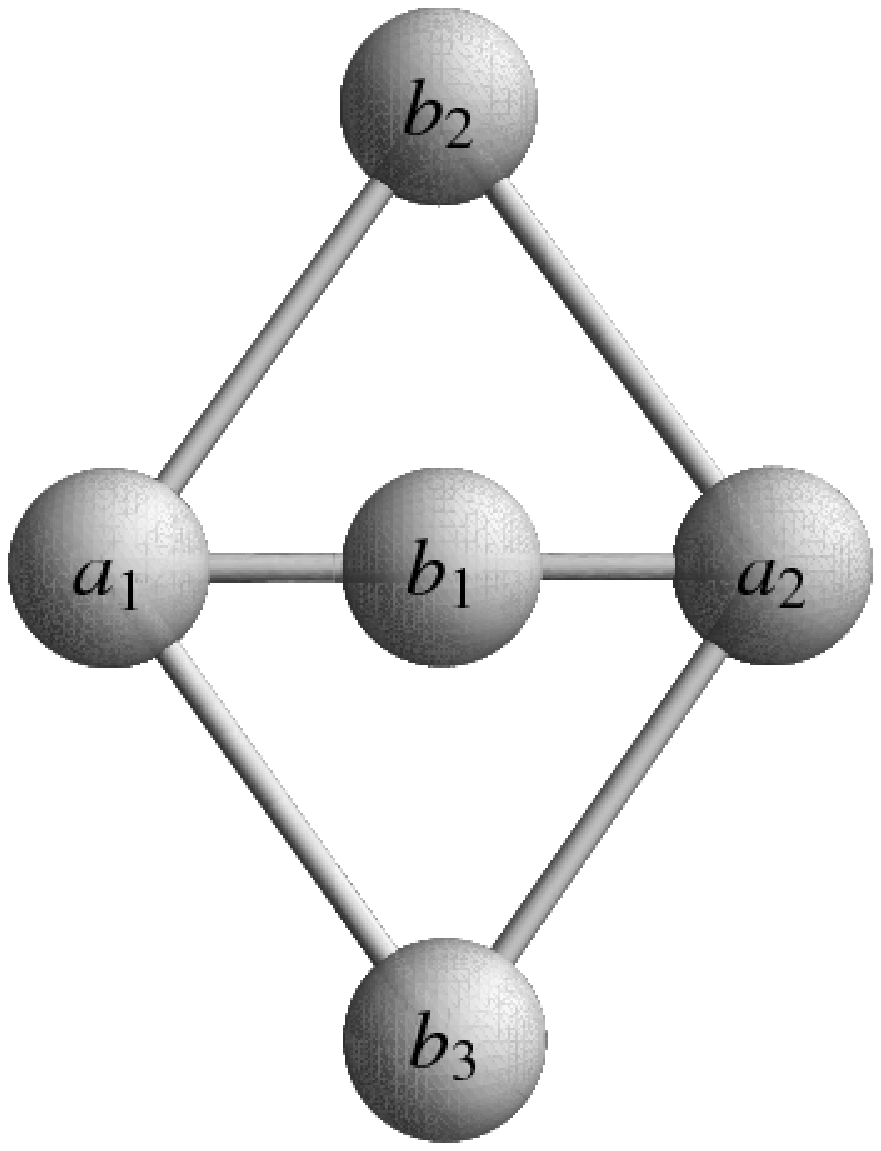} \\
$(e)$ & $(f)$ \\
\includegraphics[scale=0.4]{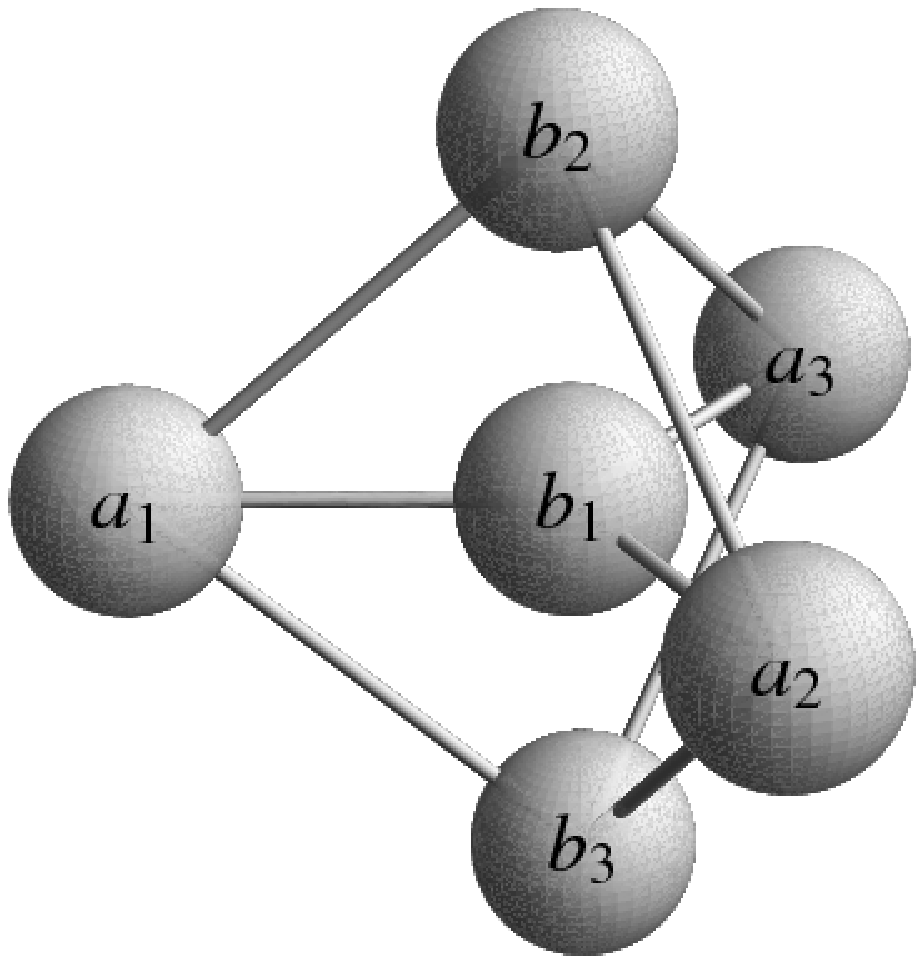} & \includegraphics[scale=0.4]{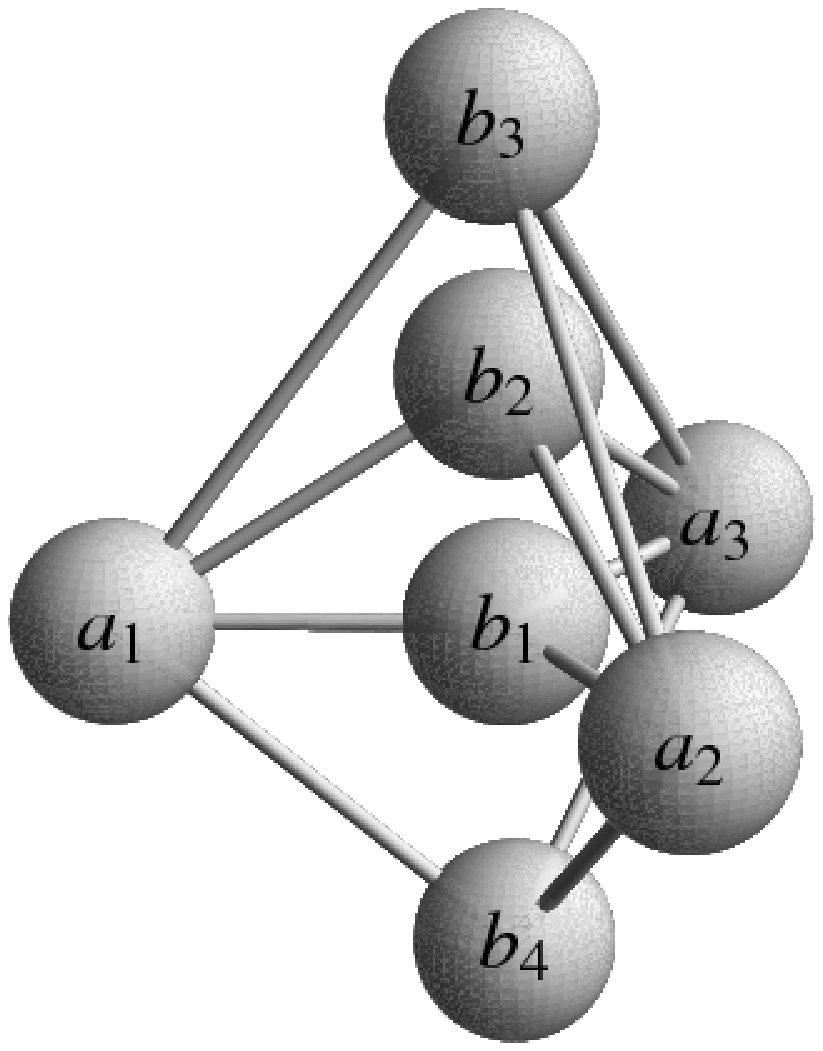} \\
\end{tabular}
\caption{In  $(a)$ we have $K_{1,1}$, in $(b)$ we have $K_{2,1}$, in $(c)$  we have $K_{2,2}$, in $(d)$ we have $K_{2,3}$, in $(e)$ we have $K_{3,3}$, in $(f)$ we have $K_{3,4}$.} 
\label{GF1}
\end{center}
\end{figure}

\begin{figure}
\begin{center}
\begin{tabular}{cc}
$(g)$ & $(h)$ \\
\includegraphics[scale=0.5]{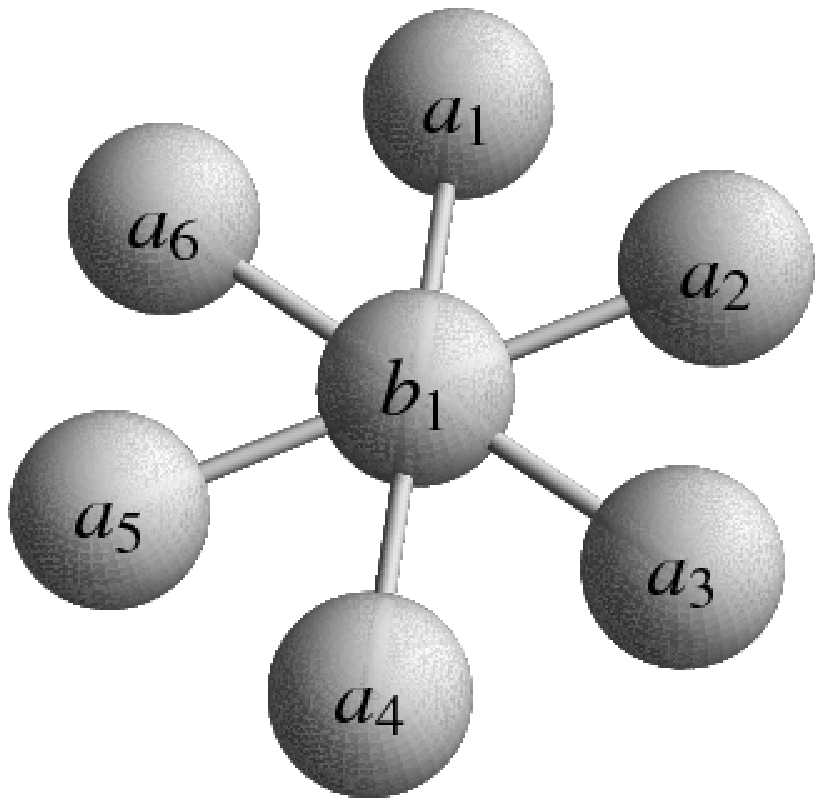} & \includegraphics[scale=0.4]{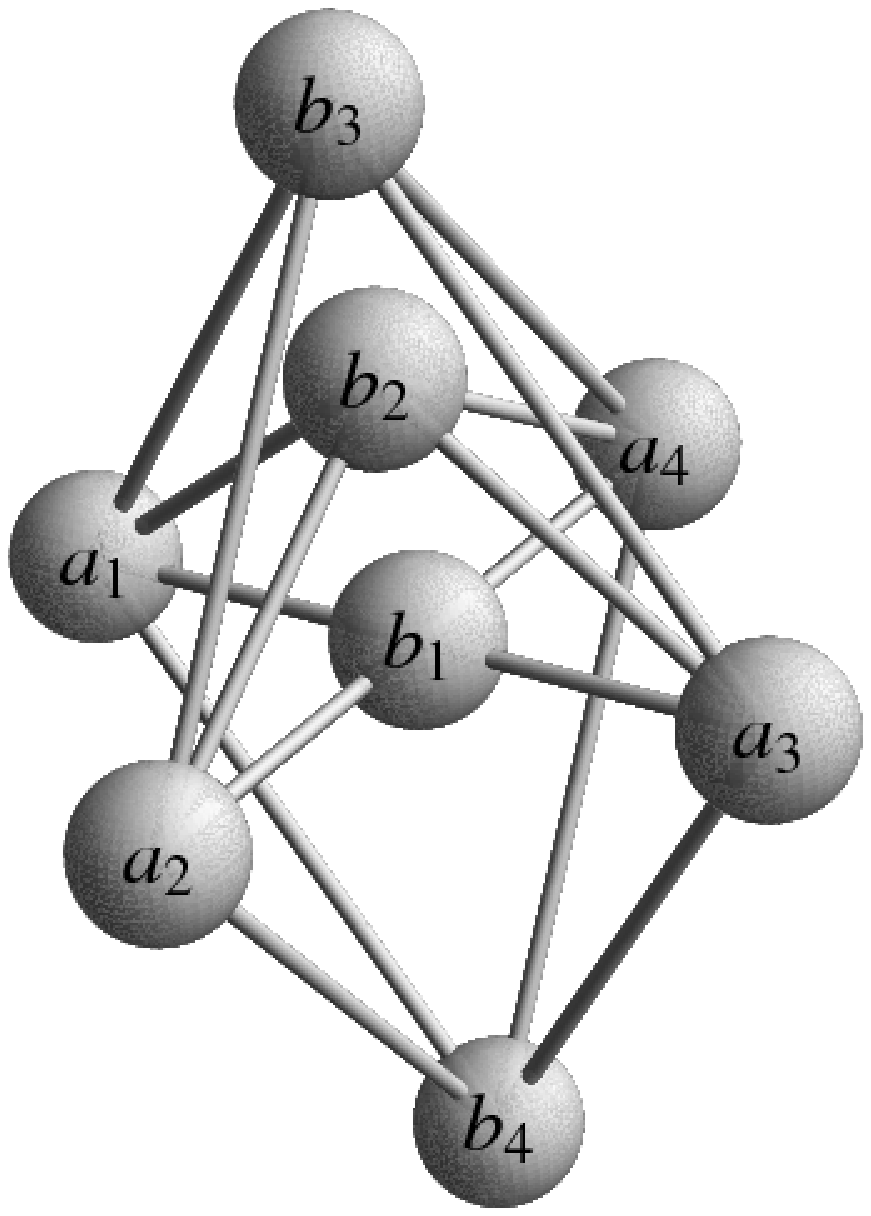} \\
$(i)$ & $(j)$ \\
\includegraphics[scale=0.4]{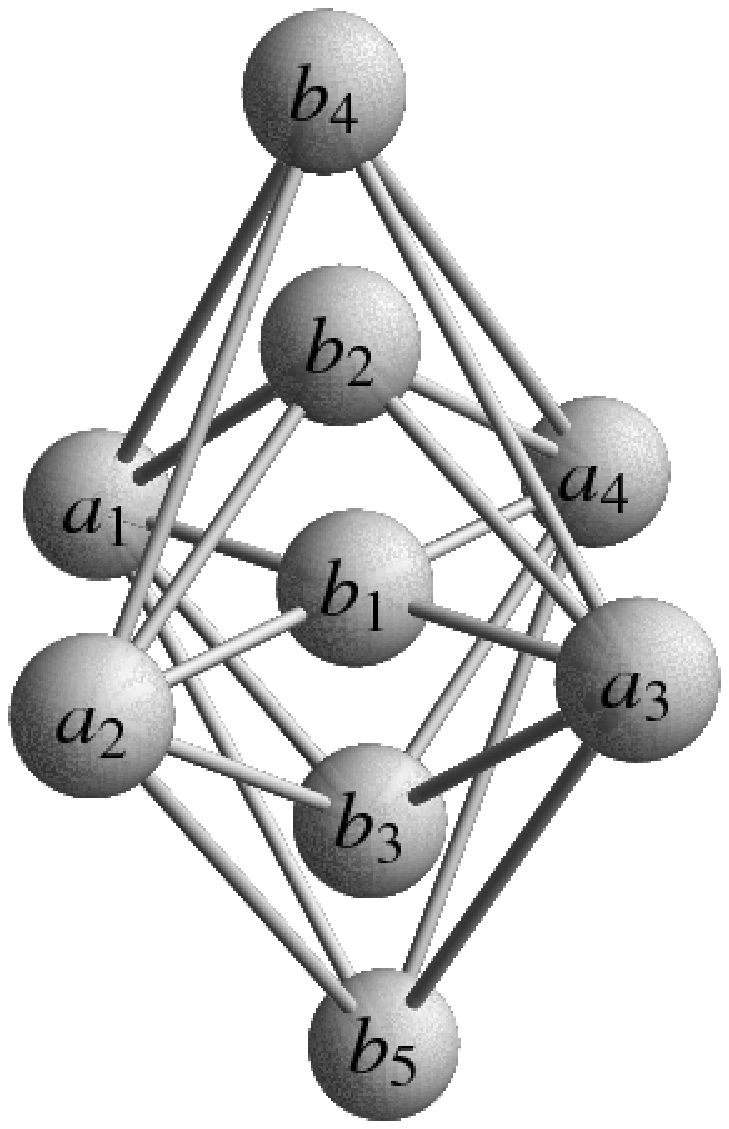} & \includegraphics[scale=0.45]{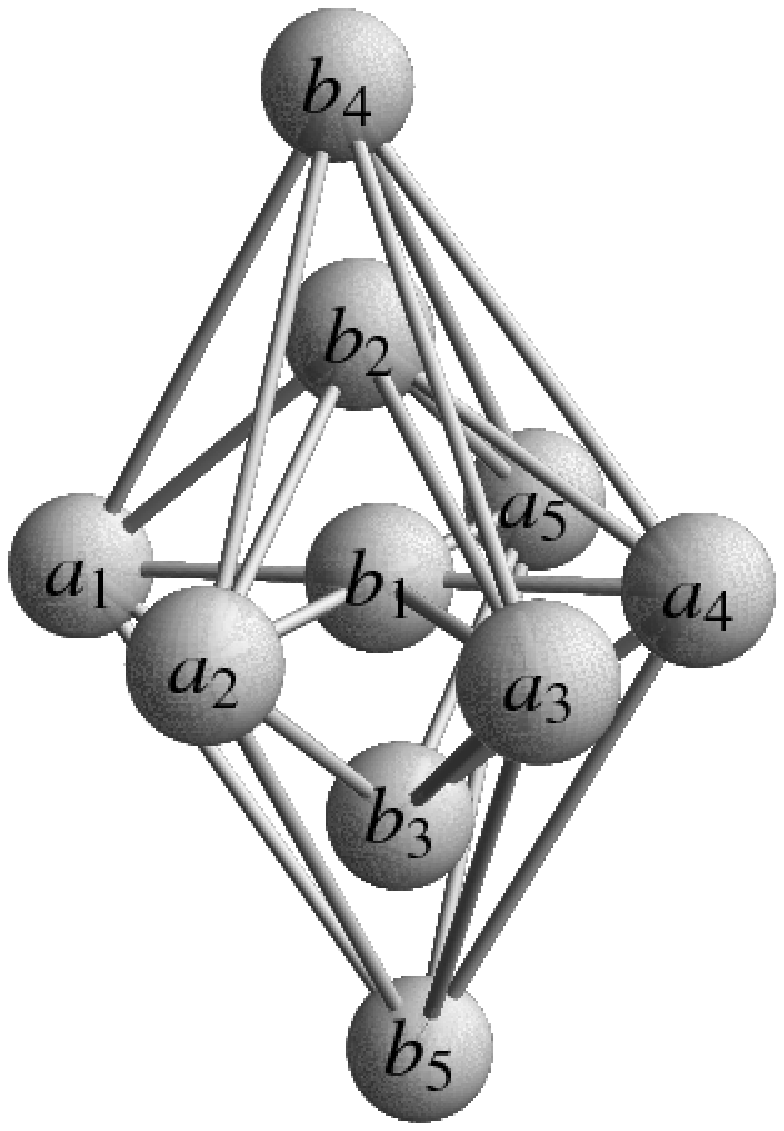} \\
\end{tabular}
\caption{In  $(g)$ we have $K_{6,1}$, in $(h)$ we have $K_{4,4}$, in $(i)$ we have $K_{4,5}$, in $(j)$ we have $K_{5,5}$.} 
\label{GF2}
\end{center}
\end{figure}

In the Figs. (\ref{GF1}) and (\ref{GF2}) we show some graphs for different values of $n$ and $m$. The balls with their respective labels are representing the condensates and the tubes are representing the tunnelling of the atoms between the respective condensates. They form the complete bipartite graph $K_{n,m}$.

\section{The twisted model}

The Hamiltonian (\ref{H1}) is invariant under a global $U(1)$ gauge transformation  for each operator $\hat{p}_{j}^{\dagger}$ and $\hat{p}_{j}$, $p= a, b$, but for a  local $U(1)$ gauge transformation

\begin{equation}
\hat{p}_{j} \rightarrow e^{i\theta_{pj}}\hat{p}_{j},  \qquad \hat{p}_{j}^{\dagger} \rightarrow \hat{p}_{j}^{\dagger}e^{-i\theta_{pj}},
\label{gtransf1}
\end{equation}
\noindent we get a twisted Hamiltonian

\begin{eqnarray}
\hat{H}_{tw} & = & \sum_{j=1}^{n} U_{ajaj}\hat{N}^2_{aj} + \sum_{k=1}^{m} U_{bkbk}\hat{N}^2_{bk}  + \frac{1}{2}\sum_{ j=1 }^{n}\sum_{ l=1 (j\neq l)}^{n} U_{ajal} \hat{N}_{aj}\hat{N}_{al} \nonumber \\ 
&+& \frac{1}{2}\sum_{ k=1 }^{m}\sum_{ s=1 (k\neq s)}^{m} U_{bkbs} \hat{N}_{bk}\hat{N}_{bs}\nonumber \\ 
&+& \sum_{j=1}^{n}\sum_{k=1}^{m} U_{ajbk} \hat{N}_{aj}\hat{N}_{bk} + \sum_{j=1}^{n} (\epsilon_{aj}  - \mu_{aj})\hat{N}_{aj} + \sum_{k=1}^{m}(\epsilon_{bk} - \mu_{bk})\hat{N}_{bk} \nonumber \\
&+&  \sum_{j=1}^{n}\sum_{k=1}^{m} J_{ajbk} (e^{-i\theta_{ajbk}}\hat{a}_{j}^{\dagger}\hat{b}_{k} + e^{i\theta_{ajbk}}\hat{b}_{k}^{\dagger}\hat{a}_{j}).
\label{H1tw1}
\end{eqnarray}
\noindent with $\theta_{ajbk} = \theta_{aj} - \theta_{bk}$.

The vectors (\ref{state1}) now becomes
\begin{eqnarray}
  |n_{a1},\ldots, n_{bm}\rangle &=& \frac{e^{-i(\sum_{j=1}^n\theta_{aj}n_{aj} + \sum_{k=1}^m\theta_{bk}n_{bk})}}{\sqrt{\prod_{j=1}^n n_{aj}!\prod_{k=1}^m n_{bk}!}}\prod_{j=1}^n (\hat{a}_{j}^{\dagger})^{n_{aj}}\prod_{k=1}^m(\hat{b}_{k}^{\dagger})^{n_{bk}}|0\rangle, \nonumber\\
\label{state2}   
\end{eqnarray}
\noindent and the wavefunction (\ref{state1g})
\begin{equation}
|\Psi\rangle = \sum_{j=1}^n\sum_{k=1}^m \frac{e^{-i(\sum_{j=1}^n\theta_{aj}n_{aj} + \sum_{k=1}^m\theta_{bk}n_{bk})}}{\sqrt{\prod_{j=1}^n n_{aj}!\prod_{k=1}^m n_{bk}!}}\prod_{j=1}^n (\hat{a}_{j}^{\dagger})^{n_{aj}}\prod_{k=1}^m(\hat{b}_{k}^{\dagger})^{n_{bk}}|0\rangle.
\label{state2g}
\end{equation}

The total number of atoms, $\hat{N} = \sum_{j=1}^n \hat{N}_{aj}  + \sum_{k=1}^m \hat{N}_{bk}$, is a conserved quantity yet, $[\hat{H}_{tw},\hat{N}] = 0$. The parity of the wavefunction,  the dimension of the space state and of the Hilbert space also don't change.

The gauge transformation (\ref{gtransf1}) changes the currents (\ref{Current-J}) and (\ref{Current-T}) 

\begin{eqnarray}
\mathcal{I}_{ajbk} &=& \frac{1}{2}(N_{aj} - N_{bk}), \\
\mathcal{J}_{ajbk} &=& \frac{1}{2i}(e^{-i\theta_{ajbk}}\hat{a}_{j}^{\dagger}\hat{b}_{k} - e^{i\theta_{ajbk}}\hat{b}_{k}^{\dagger}\hat{a}_{j}),  \\
\mathcal{T}_{ajbk} &=& \frac{1}{2}(e^{-i\theta_{ajbk}}\hat{a}_{j}^{\dagger}\hat{b}_{k} + e^{i\theta_{ajbk}}\hat{b}_{k}^{\dagger}\hat{a}_{j}). 
\label{Currents-gt} 
\end{eqnarray}
\noindent but preserves the currents algebra (\ref{CA}).

The gauge parameters $\theta_{pj}$ can be chooses arbitrarily. The difference between them $\theta_{ajbk}$ can be considered, as for example, a magnetic flux to study Meissner effect and Aharonov-Bohm in a loop of BECs \cite{ladder1,ladder2,goldman,goldman2,orignac,haller2,galitski} or a  twist angle, by Peierls phase factor, when we consider superfluid fraction in the BECs \cite{roth,jardel1,jardel2}.

\section{Exact solution of the fragmented model}

In this section we present the exact solution of the BEC fragmented model using the Lax operators (\ref{L2}) and (\ref{L3}) for models with different number of BECs $a$ and $b$.

 Using the co-multiplication property of the Lax operators we can write,

\begin{equation}
  \hat{L}(u) = \hat{L}_{1}^{\Sigma_a^n}(u + \sum_{j=1}^{n}\omega_{aj})\hat{L}_{2}^{\Sigma_b^m}(u - \sum_{k=1}^{m}\omega_{bk}).
\label{LH2}
\end{equation}
\noindent Following the monodromy matrix (\ref{monod}) we can write the operators,

\begin{eqnarray}
\pi(\hat{A}(u)) &=& \left(u\hat{I} + \hat{I}\sum_{j=1}^n \omega_{aj} + \eta\sum_{j=1}^n \hat{N}_{aj}\right)\left(u\hat{I}  - \hat{I}\sum_{k=1}^m \omega_{bk} + \eta\sum_{k=1}^m \hat{N}_{bk}\right) \nonumber \\
          &+& \sum_{j=1}^n\sum_{k=1}^m t_{aj}s_{bk} \hat{b}_k^{\dagger}\hat{a}_j, \label{piAn}\\
\pi(\hat{B}(u)) &=& \left(u\hat{I} + \hat{I}\sum_{j=1}^n \omega_{aj} + \eta\sum_{j=1}^n \hat{N}_{aj}\right)\left(\sum_{k=1}^m t_{bk}\hat{b}_k\right) 
+ \frac{\zeta_b}{\eta}\sum_{j=1}^n t_{aj}\hat{a}_j, \label{piBn}\\
\pi(\hat{C}(u)) &=& \left(\sum_{j=1}^n s_{aj}\hat{a}_j^{\dagger}\right)\left(u\hat{I}  - \hat{I}\sum_{k=1}^m \omega_{bk} + \eta\sum_{k=1}^m \hat{N}_{bk}\right) 
          + \frac{\zeta_a}{\eta}\sum_{k=1}^m s_{bk}\hat{b}_k^{\dagger}, \label{piCn}\\
\pi(\hat{D}(u)) &=& \sum_{j=1}^n\sum_{k=1}^m s_{aj}t_{bk}\hat{a}_j^{\dagger}\hat{b}_k + \eta^{-2}\zeta_a\zeta_b \hat{I}. \label{piDn}
\end{eqnarray}

Taking the trace of the operator (\ref{LH2}) we get the transfer matrix

\begin{eqnarray}
\hat{t}(u) & = & u^2 \hat{I} + u\hat{I}\left(\sum_{j=1}^n \omega_{aj}-\sum_{k=1}^m \omega_{bk}\right) + u\eta\hat{N} \nonumber\\ 
&+&  \left(\frac{\zeta_a\zeta_b}{\eta^{2}} - \sum_{j=1}^n \sum_{k=1}^m \omega_{aj}\omega_{bk}\right)\hat{I}  \nonumber\\
	 & - & \eta\sum_{j=1}^n\sum_{k=1}^m \left(\omega_{bk}\hat{N}_{aj} -  \omega_{aj}\hat{N}_{bk}   \right)  + \eta^2\sum_{j=1}^n \sum_{k=1}^m \hat{N}_{aj}\hat{N}_{bk}  \nonumber \\
 	 & + & \sum_{j=1}^{n}\sum_{k=1}^{m}(s_{aj}t_{bk}\hat{a}_{j}^{\dagger}\hat{b}_{k} + s_{bk}t_{aj}\hat{b}_{k}^{\dagger}\hat{a}_{j}).
\label{tu2} 
\end{eqnarray}

\noindent From (\ref{C14b}) we identify the conserved quantities of the transfer matrix (\ref{tu2}),

\begin{eqnarray}
\hat{\mathcal{C}}_0 
	 & = &  \left(\eta^{-2}\zeta_a\zeta_b - \sum_{j=1}^n \sum_{k=1}^m \omega_{aj}\omega_{bk}\right)\hat{I}  \nonumber\\
	 & - & \eta\sum_{j=1}^n\sum_{k=1}^m \left(\omega_{bk}\hat{N}_{aj} -  \omega_{aj}\hat{N}_{bk}   \right)  + \eta^2\sum_{j=1}^n \sum_{k=1}^m \hat{N}_{aj}\hat{N}_{bk}  \nonumber \\
 	 & + & \sum_{j=1}^{n}\sum_{k=1}^{m}(s_{aj}t_{bk}\hat{a}_{j}^{\dagger}\hat{b}_{k} + s_{bk}t_{aj}\hat{b}_{k}^{\dagger}\hat{a}_{j}),
\label{cquantity0}
\end{eqnarray}

\begin{eqnarray}
\hat{\mathcal{C}}_1 &=&  \hat{I}\left(\sum_{j=1}^n \omega_{aj}-\sum_{k=1}^m \omega_{bk}\right) + \eta\hat{N},
\label{cquantity1}
\end{eqnarray}

\begin{eqnarray}
\hat{\mathcal{C}}_2 &=& \hat{I}.
\label{cquantity2}
\end{eqnarray}
We can rewrite the Hamiltonian (\ref{H1}) using these conserved quantities

\begin{equation}
 \hat{H} = \xi_0\hat{\mathcal{C}}_0  + \xi_1\hat{\mathcal{C}}_1^2 + \xi_2\hat{\mathcal{C}}_2,
\label{H4} 
\end{equation}
with the following identification for the parameters
\begin{eqnarray}
\xi_2 &=& - \xi_0\left( \eta^{-2} \zeta_a\zeta_b - \omega_{ab}^{nm}\right) - \xi_1\Delta\omega_{ab}^2,\\
\Delta\omega_{ab} &=& \sum_{j=1}^n \omega_{aj} - \sum_{k=1}^m \omega_{bk}, \\
\omega_{ab}^{nm}  &=& \sum_{j=1}^n\sum_{k=1}^m \omega_{aj}\omega_{bk},
\end{eqnarray}

\begin{equation}
U_{ajaj} = U_{bkbk} = \eta^2\xi_1, \qquad U_{ajal} =U_{bkbs} = 2\eta^2\xi_1,
\end{equation}

\begin{equation}
U_{ajbk} = \eta^2(\xi_0 + 2\xi_1), 
\end{equation}

\begin{equation}
\epsilon_{aj} - \mu_{aj} = 2\xi_1\eta\Delta\omega_{ab} - \xi_0\eta\sum_{k=1}^m \omega_{bk},
\end{equation}

\begin{equation}
\epsilon_{bk} - \mu_{bk} = 2\xi_1\eta\Delta\omega_{ab} + \xi_0\eta\sum_{j=1}^n \omega_{aj},
\end{equation}

\begin{equation}
J_{ajbk} = -\xi_0 t_{aj} s_{bk} = -\xi_0 s_{aj} t_{bk}.
\end{equation}

The Hamiltonians (\ref{H1}) and (\ref{H4}) are related with the transfer matrix (\ref{tu2}) by the equation,

\begin{equation}
 \hat{H} = \xi_0\hat{t}(u) + \xi_1\hat{\mathcal{C}}_1^2 - \xi_0\hat{\mathcal{C}}_1u - (\xi_0 u^2 - \xi_2)\hat{\mathcal{C}}_2.
\label{H5} 
\end{equation}
\noindent In the Hamiltonian (\ref{H5}) the spectral parameter $u$ is canceled for any value and we get again the Hamiltonian (\ref{H4}).

We use as pseudovacuum the product state, 

\begin{equation}
|0\rangle = \left(\bigotimes_{j=1}^n|0\rangle_{a_j}\right)\otimes \left(\bigotimes_{k=1}^m|0\rangle_{b_k}\right),
\label{pvacuum}
\end{equation}
\noindent with $|0\rangle_{a_j}$ denoting the Fock vacuum state for the BECs $a_j$ and $|0\rangle_{b_k}$ denoting the Fock vacuum state for the BECs $b_k$, for $j=1,\ldots, n$ and $k=1\ldots, m$. For this pseudo-vacuum we can apply the ABA method in order to find the Bethe ansatz equations (BAEs),

\begin{eqnarray}
\frac{v^2_{i} + v_{i}\Delta\omega_{ab} - \omega_{ab}^{nm}}{\eta^{-2}\zeta_a\zeta_b} & = & \prod_{j \ne i}^{N}\frac{v_{i}-v_{j}-\eta}{v_{i}-v_{j}+\eta}, \;\;\;\;\;  i,j = 1,\ldots , N. \nonumber\\
\label{BAE2}
\end{eqnarray}

The eigenvectors \cite{Bethe-states}  $\{ |v_1,\ldots,v_N\rangle \}$ of the Hamiltonian (\ref{H1}) or (\ref{H4}) and of the transfer matrix (\ref{tu2}) are 

\begin{equation}
|\vec{v}\rangle = \prod_{i=1}^N \left[\left(\sum_{j=1}^n s_{aj}\hat{a}_j^{\dagger}\right)\left(v_i - \sum_{k=1}^m \omega_{bk} + \eta\sum_{k=1}^m \hat{N}_{bk}\right) 
          + \frac{\zeta_a}{\eta}\sum_{k=1}^m s_{bk}\hat{b}_k^{\dagger}\right]|0 \rangle,
\end{equation}
\noindent and the eigenvalues of the Hamiltonian (\ref{H1}) or (\ref{H4}) are,

\begin{eqnarray}
E(\{ v_i \}) & = &   \xi_0\left(u^2 + u\Delta\omega_{ab} - \omega_{ab}^{nm}\right) \prod_{i=1}^{N}\frac{v_{i} - u +\eta}{v_{i} - u} \nonumber \\
  & + &  \xi_0\eta^{-2}\zeta_a\zeta_b\prod_{i=1}^{N}\frac{v_{i} - u -\eta}{v_{i} - u} + \xi_1\mathcal{C}_1^2 - \xi_0\mathcal{C}_1u - \xi_0 u^2 + \xi_2,
\label{eigenH1-ia}  
\end{eqnarray}
\noindent where the $\{v_j\}$ are solutions of the BAEs (\ref{BAE2}) and $N$ is the total number of atoms. Because  (\ref{H5}), we can choose the spectral parameter $u$ arbitrarily. 

Choosing $\frac{J_{ajbk}}{\xi_0} = \pm \eta$, for example, we can write the BAEs (\ref{BAE2}) in the no interaction  limit, $\eta \rightarrow 0$, as  just one equation
\begin{eqnarray}
\sum_{i=1}^N \left(v_{i} + \frac{1}{2}\Delta\omega_{ab} \right)^2  & = & R_N^2,  \nonumber\\
\label{BAE3}
\end{eqnarray}  
\noindent with

\begin{equation}
R_N = \sqrt{\left(\frac{1}{4}\Delta\omega^2_{ab} + \omega_{ab}^{nm} + n\times m \right) N}.
\end{equation}
\noindent If the Bethe roots $\{v_i\}$ are real numbers, the BAE (\ref{BAE3}) is the equation of a $N$-dimensional sphere of radius $R_N$ and center in 

\begin{equation}
v_{i} = - \frac{1}{2}\Delta\omega_{ab}, \qquad \forall \; i = 1, \ldots, N.
\end{equation}

If the Bethe roots $\{v_i\}$ are complex numbers, the complex conjugate of them, $\{\bar{v}_i\}$, are also solutions of the equation (\ref{BAE3}) and we can write this equation as

\begin{equation}
\sum_{i=1}^N \frac{\left(x_{i} + \frac{1}{2}\Delta\omega_{ab} \right)^2}{R_N^2} - \sum_{i=1}^N \frac{y_{i}^2}{R_N^2} = 1.
\label{pcs}
\end{equation}

 The Eq. (\ref{pcs}) is the equation of a proper central surface with semiaxes $R_N$ in a  $\mathbb{R}^{2N+1}$ space. From Eq. (\ref{BAE3}) we can see that there are $N!$ permutations for each set of Bethe roots $\{v_i\}$. 

In the limit $\eta \rightarrow 0$ and with $u = 0$ we can write the eigenvalues as

\begin{eqnarray}
E(\{ v_i \}) & = &   \xi_1\left(\mathcal{C}_1^2  - \Delta\omega_{ab}^{2}\right)  -  \eta\xi_0\left(n\times m + \omega_{ab}^{nm}\right)\sum_{i=1}^{N} \frac{1}{v_{i}}.
\label{eigenH1-nib}  
\end{eqnarray}

\section{Exact solution of the twisted model}

In this section we will show that the twisted Hamiltonian (\ref{H1tw1}) has the same spectrum of the Hamiltonian (\ref{H1}) for the BECs fragmented model.

Using the new definition for the parameters $t_{pr}$ and $s_{pr}$, $p=a,b$ and $r=j,k$,

\begin{equation}
t_{pr} \rightarrow \rho_{pr} e^{i\theta_{pr}}, \qquad s_{pr} \rightarrow \kappa_{pr} e^{-i\theta_{pr}},
\label{transf2}
\end{equation}
\noindent we write the transfer matrix (\ref{tu2}) as

\begin{eqnarray}
\hat{t}(u) & = & u^2 \hat{I} + u\hat{I}\left(\sum_{j=1}^n \omega_{aj}-\sum_{k=1}^m \omega_{bk}\right) + u\eta\hat{N} \nonumber\\ 
&+&  \left(\frac{\zeta_a\zeta_b}{\eta^{2}} - \sum_{j=1}^n \sum_{k=1}^m \omega_{aj}\omega_{bk}\right)\hat{I}  \nonumber\\
	 & - & \eta\sum_{j=1}^n\sum_{k=1}^m \left( \omega_{bk}\hat{N}_{aj} -  \omega_{aj}\hat{N}_{bk}   \right)  + \eta^2\sum_{j=1}^n \sum_{k=1}^m \hat{N}_{aj}\hat{N}_{bk}  \nonumber \\
 	 & + & \sum_{j=1}^{n}\sum_{k=1}^{m}(\kappa_{aj}\rho_{bk}e^{-i\theta_{ajbk}}\hat{a}_{j}^{\dagger}\hat{b}_{k} + \rho_{aj}\kappa_{bk}e^{i\theta_{ajbk}}\hat{b}_{k}^{\dagger}\hat{a}_{j}).
\label{tu3} 
\end{eqnarray}
\noindent  if the conditions, $\zeta_a = \sum_{j=1}^{n} \rho_{aj}\kappa_{aj}$ and $\zeta_b = \sum_{k=1}^{m}\rho_{bk}\kappa_{bk}$, are satisfied. The reparametrization (\ref{transf2}) is equivalent to the $U(1)$ local gauge transformation (\ref{gtransf1}).

From (\ref{C14b}) we identify the conserved quantities of the transfer matrix (\ref{tu3}),

\begin{eqnarray}
\hat{\mathcal{C}}_0 
	 & = &  \left(\eta^{-2}\zeta_a\zeta_b - \sum_{j=1}^n \sum_{k=1}^m \omega_{aj}\omega_{bk}\right)\hat{I}  \nonumber\\
	 & - & \eta\sum_{j=1}^n\sum_{k=1}^m \left( \omega_{bk}\hat{N}_{aj} -  \omega_{aj}\hat{N}_{bk}   \right)  + \eta^2\sum_{j=1}^n \sum_{k=1}^m \hat{N}_{aj}\hat{N}_{bk}  \nonumber \\
 	 & + & \sum_{j=1}^{n}\sum_{k=1}^{m}(\kappa_{aj}\rho_{bk}e^{-i\theta_{ajbk}}\hat{a}_{j}^{\dagger}\hat{b}_{k} + \rho_{aj}\kappa_{bk}e^{i\theta_{ajbk}}\hat{b}_{k}^{\dagger}\hat{a}_{j}),
\end{eqnarray}

\begin{eqnarray}
\hat{\mathcal{C}}_1 &=&  \hat{I}\left(\sum_{j=1}^n \omega_{aj}-\sum_{k=1}^m \omega_{bk}\right) + \eta\hat{N},
\end{eqnarray}

\begin{eqnarray}
\hat{\mathcal{C}}_2 &=& \hat{I}.
\end{eqnarray}
We can rewrite the Hamiltonian (\ref{H1tw1}) using these conserved quantities

\begin{equation}
 \hat{H} = \xi_0\hat{\mathcal{C}}_0  + \xi_1\hat{\mathcal{C}}_1^2 + \xi_2\hat{\mathcal{C}}_2,
\label{H1tw2} 
\end{equation}
with the following identification for the parameters
\begin{eqnarray}
\xi_2 &=& - \xi_0\left( \eta^{-2} \zeta_a\zeta_b - \omega_{ab}^{nm}\right) - \xi_1\Delta\omega_{ab}^2,\\
\Delta\omega_{ab} &=& \sum_{j=1}^n \omega_{aj} - \sum_{k=1}^m \omega_{bk}, \\
\omega_{ab}^{nm}  &=& \sum_{j=1}^n\sum_{k=1}^m \omega_{aj}\omega_{bk},
\end{eqnarray}

\begin{equation}
U_{ajaj} = U_{bkbk} = \eta^2\xi_1, \qquad U_{ajal} =U_{bkbs} = 2\eta^2\xi_1,
\end{equation}

\begin{equation}
U_{ajbk} = \eta^2(\xi_0 + 2\xi_1), 
\end{equation}

\begin{equation}
\epsilon_{aj} - \mu_{aj} = 2\xi_1\eta\Delta\omega_{ab} - \xi_0\eta\sum_{k=1}^m \omega_{bk},
\end{equation}

\begin{equation}
\epsilon_{bk} - \mu_{bk} = 2\xi_1\eta\Delta\omega_{ab} + \xi_0\eta\sum_{j=1}^n \omega_{aj},
\end{equation}

\begin{equation}
J_{ajbk} = -\xi_0 \rho_{aj} \kappa_{bk} = -\xi_0  \rho_{bk}\kappa_{aj}.
\end{equation}

The Hamiltonian (\ref{H1tw1}) is related with the transfer matrix (\ref{tu3}) by the equation,

\begin{equation}
 \hat{H} = \xi_0\hat{t}(u) + \xi_1\hat{\mathcal{C}}_1^2 - \xi_0\hat{\mathcal{C}}_1u - (\xi_0 u^2 - \xi_2)\hat{\mathcal{C}}_2.
\label{H1tw3} 
\end{equation}

Using the same pseudo-vacuum product state we find the same Bethe ansatz equations,

\begin{eqnarray}
\frac{v^2_{i} + v_{i}\Delta\omega_{ab} - \omega_{ab}^{nm}}{\eta^{-2}\zeta_a\zeta_b} & = & \prod_{j \ne i}^{N}\frac{v_{i}-v_{j}-\eta}{v_{i}-v_{j}+\eta}, \;\;\;\;\;  i,j = 1,\ldots , N, \nonumber\\
\label{BAE2tw}
\end{eqnarray}
\noindent that are independent of the twist angles and if we make the choice for the parameters  
\begin{equation}
\zeta_a = \sum_{j=1}^{n} \rho_{aj}\kappa_{aj} = \sum_{j=1}^{n}s_{aj}t_{aj}, \qquad \zeta_b = \sum_{k=1}^{m}\rho_{bk}\kappa_{bk} = \sum_{k=1}^{m}s_{bk}t_{bk}.
\end{equation}

Therefore, the eigenvectors \cite{Bethe-states}  $\{ |v_1,\ldots,v_N\rangle \}$ now are different because of the gauge parameters 

\begin{eqnarray}
|\vec{v}\rangle &=& \prod_{i=1}^N \left[\left(\sum_{j=1}^n \kappa_{aj} e^{-i\theta_{aj}}\hat{a}_j^{\dagger}\right)\left(v_i - \sum_{k=1}^m \omega_{bk} + \eta\sum_{k=1}^m \hat{N}_{bk}\right)\right. 
       \nonumber \\   
       &+& \left. \frac{\zeta_a}{\eta}\sum_{k=1}^m \kappa_{bk} e^{-i\theta_{bk}}\hat{b}_k^{\dagger}\right]|0 \rangle,
\label{BV2}
\end{eqnarray}
\noindent but the eigenvalues of the Hamiltonian (\ref{H1tw1}) or (\ref{H1tw3}) are the same of the Hamiltonian (\ref{H1}),

\begin{eqnarray}
E(\{ v_i \}) & = &   \xi_0\left(u^2 + u\Delta\omega_{ab} - \omega_{ab}^{nm}\right) \prod_{i=1}^{N}\frac{v_{i} - u +\eta}{v_{i} - u} \nonumber \\
  & + &  \xi_0\eta^{-2}\zeta_a\zeta_b\prod_{i=1}^{N}\frac{v_{i} - u -\eta}{v_{i} - u} + \xi_1\mathcal{C}_1^2 - \xi_0\mathcal{C}_1u - \xi_0 u^2 + \xi_2,
\label{eigenHtw-i}  
\end{eqnarray}
\noindent where the $\{v_i\}$ are solutions of the BAEs (\ref{BAE2tw}) and $N$ is the total number of atoms. We can choose arbitrarily the spectral parameter $u$. In the no interaction limit, $\eta \rightarrow 0$ , and $u=0$ we get the same equation (\ref{eigenH1-nib}).

\section{Summary}

We have solved a family of fragmented Bose-Hubbard models using the multi-states  boson Lax operator \cite{multistates}. These models can be considered as graphs, with the BECs in the vertices and the edge representing the tunnelling between the BECs. The graphs form the complete bipartite graph $K_{n,m}$. We have showed that in the no interaction limit, $\eta \rightarrow 0$, if the Bethe roots are all real numbers, they are on a $N$-dimensional sphere of radius $R_N$. If the Bethe roots are complex numbers, they are on a $2N$-dimensional proper central surface with semiaxes $R_N$. For each Bethe roots set we have a set of $N!$ permutations of the Bethe roots that satisfy the Bethe ansatz equation. By a local $U(1)$ gauge transformation we got twisted models and showed that they have the same spectrum of eigenvalues but with different wavefunctions. The new wavefunction presents a pattern of interference dependent of the gauge transformation parameters and preserve the parity.

\section*{Acknowledgments}
The author acknowledge Capes/FAPERJ (Coordena\c{c}\~ao de Aperfei\c{c}oamento de Pessoal de N\'{\i}vel Superior/Funda\c{c}\~ao de Amparo \`a Pesquisa do Estado do Rio de Janeiro) for the initial financial support. The author also acknowledge the physics department of the Universidade Federal de Sergipe UFS for support.


\section*{References}
\begin{thebibliography}{10}

\bibitem{bose} Bose S N, \textit{Z. Phys.}  \textbf{26} (1924) 178.

\bibitem{einstein} Einstein A, \textit{Phys. Math. K1}  \textbf{22} (1924) 261.
\bibitem{anderson} Anderson M H, Ensher J R, Mathews M R, Wieman C E and Cornell E A, \textit{Science}  {\bf 269} (1995) 198.

\bibitem{Hulet} Sackett C A,  Bradley C C, Welling M and Hulet R G, \textit{Braz.  Jour. Phys.} \textbf{27} no. 2 (1997) 154.

\bibitem{wwcch} Williams J, Walser R, Cooper J, Cornell E A and Holland M, \textit{Phys. Rev. A} \textbf{61} (2000) 0336123.

\bibitem{early} Cornell E A  and Wieman C E, \textit{Rev. Mod. Phys.} \textbf{74} (2002) 875.

\bibitem{wk} Ketterle W, \textit{Rev. Mod. Phys.} \textbf{74} (2002) 1131.

\bibitem{angly} Anglin J R  and  Ketterle W, \textit{Nature} \textbf{416} (2002) 211.

\bibitem{Bagnato} Mancini M.W., Telles G. D., Caires  A. R. L., Bagnato V. S.  and Marcassa L. G. , \textit{Phys. Rev. Lett.}  \textbf{92}  (2004) 133203.

\bibitem{Yukalov} Courteille Ph. W.,  Bagnato V. S. and Yukalov V. I. \textit{Laser Physics} \textbf{11}, No. 6, 2001, pp. 659.

\bibitem{Josephson1} Josephson B D,  \textit{Phys. Lett.} \textbf{1} (1962) 251.

\bibitem{Josephson2} Josephson B D,  \textit{Rev. Mod. Phys.} \textbf{46} (1974) 251.

\bibitem{Ketterle1} Andrews M R, Townsend C G, Miesner H-J, Durfee D S, Kurn D M  and Ketterle W,  \textit{Science} \textbf{275} (1997) 637. 

\bibitem{Ketterle2} Shin Y, Saba M, Pasquini T A, Ketterle W, Pritchard D E  and Leanhardt A E,  \textit{Phys. Rev. Lett.} \textbf{92} (2004) 050405.

\bibitem{Oberthaler1} Albiez M, Gati R, F\"olling J, Hunsmann S,   
Cristiani M and  Oberthaler M K, \textit{Phys. Rev. Lett.} \textbf{95} (2005) 010402.



\bibitem{Oberthaler2} Zibold T, Nicklas  E, Gross C and Oberthaler M K,  \textit{Phys. Rev. Lett.} \textbf{105} (2010) 204101.

\bibitem{Oberthaler3} Gati R, Albiez M, F\"olling J, Hemmerling B and Oberthaler M K, \textit{Appl. Phys.} \textbf{B 82} (2006) 207.  
\bibitem{Bloch} Bloch I, Dalibard J and Zwerger W, \textit{Rev. Mod. Phys.}  \textbf{80} (2008) 875.
\bibitem{Ryu1} Henderson K, Ryu C, MacCormick C and Boshier M G, \textit{New J. Phys.} \textbf{11} (2009) 043030.

\bibitem{Ryu2} Ryu C, Blackburn P W, Blinova A A and Boshier M G, \textit{Phys. Rev. Lett.} \textbf{111}  (2013) 205301.

\bibitem{ladder1} Atala M, Aidelsburger M, Lohse M, Barreiro J T, Paredes B and Bloch I, \textit{Nat. Phys.} \textbf{10} (2014) 588.

\bibitem{ladder2} Kolley F, Piraud M, McCulloch I P, Schollw\"ock U  and Heidrich-Meisner F, \textit{New J. Phys.} \textbf{17} (2015) 092001.

\bibitem{goldman} N Goldman, G Juzeliunas, P \"Ohberg and I B Spielman, \textit{Rep. Prog. Phys.}  \textbf{77} (2014) 126401. 

\bibitem{goldman2} M Aidelsburger, S Nascimbene, N Goldman, \textit{C. R. Physique} \textbf{19} (2018) 394.

\bibitem{orignac} E Orignac, R Citro, M Di Dio, S De Palo and M-L Chiofalo, \textit{New J. Phys.} \textbf{18} (2016) 055017.

\bibitem{haller2} A Haller, A S. Matsoukas-Roubeas, Y. Pan, M Rizzi and M Burrello, \textit{Phy. Rev.  Res.}  \textbf{2}  (2020) 043433.

\bibitem{galitski} V Galitski, I Spielman and G Juzeliunas, \textit{Physics Taday} \textbf{72(1)} (2019) 39. 

\bibitem{Roditi} Roditi I, \textit{Brazilian Journal of Physics} \textbf{30} (2000) 357.

\bibitem{jon1} Zhou H-Q, Links J, Gould M and McKenzie R, \textit{J. Math. Phys.} \textbf{44} (2003) 4690.


\bibitem{jon2} Zhou H-Q, Links J and McKenzie R H, \textit{Int. Jour. Mod. Phys. B} \textbf{17} (2003) 5819.

\bibitem{jonjpa} Links J, Zhou H-Q, McKenzie R H  and Gould M D, \textit{J. Phys. A} \textbf{36} (2003) R63.

\bibitem{key-3} Foerster A , Links J and Zhou H-Q, \textit{Classical and quantum non-linear integrable systems: theory and applications}, Editor A. Kundu, IOP Publishing, Bristol and Philadelphia, (2003) 208.



\bibitem{dukelskyy} Dukelsky J, Dussel G, Esebbag C  and Pittel S, \textit{Phys. Rev. Lett.} \textbf{93} (2004) 050403.

\bibitem{Ortiz} Ortiz G, Somma R, Dukelsky J and Rombouls S, \textit{Nuclear Physics B} \textbf{707} (2005) 421.
 
\bibitem{Kundu} Kundu A, \textit{Theoretical and Mathematical Physics} \textbf{151} (2007) 831.

\bibitem{eric5} Foerster A  and Ragoucy E, \textit{Nuclear Physics B} \textbf{777} (2007) 373.

\bibitem{GSantosaa} Links J, Foerster A, Tonel A P and Santos G, \textit{Ann. Henri Poincar\'e} \textbf{7} (2006) 1591.


\bibitem{GSantos} Santos G, Foerster A, Roditi I, Santos Z V T and Tonel A P, \textit{J. Phys. A: Math. Theor.} \textbf{41} (2008) 295003.


\bibitem{GSantos11} Santos G, \textit{J. Phys. A: Math. Theor.} \textbf{44} (2011) 345003.

\bibitem{rubeni} Rubeni D, Foerster A, Mattei E, and Roditi I, \textit{Nuclear Physics B} \textbf{856} (2012) 698.

\bibitem{multistates} Santos G, Foerster A  and Roditi I, \textit{J. Phys. A: Math. Theor.} \textbf{46} (2013) 265206 (12pp).

\bibitem{Xin} Zhi-Rong X, Tao Y, Kun H and Wen-Li Y, \textit{Commun. Theor. Phys.} \textbf{64} (2015) 653.

\bibitem{Ymai} Tonel A P, Ymai L H, Foerster A  and Links J, \textit{J. Phys. A: Math. Theor.} \textbf{48} (2015) 494001 (12pp).

\bibitem{Bethe-states} Santos G, Ahn C, Foerster A and Roditi I, \textit{Phys. Lett. B} \textbf{746} (2015) 186.

\bibitem{batchelor2} Murray T B and Foerster A,  \textit{J. Phys. A: Math. Theor.} \textbf{49} (2016) 173001.



\bibitem{screp} Vind F A, Foerster A, Oliveira  I S, Sarthour R S, Soares-Pinto D O, Souza A M and 
 Roditi I, \textit{Nature: Scientific Rep.} \textbf{6} (2016) 1.


\bibitem{kino1} Kinoshita T,  Wenger T and  Weiss D S, \textit{Science} \textbf{305} (2004) 1125.

\bibitem{kino2} Kinoshita T,  Wenger T and  Weiss D S, \textit{Nature} \textbf{440} (2006) 900.

\bibitem{kitagawa} Kitagawa T, Pielawa S, Imambekov A, Schmiedmayer J, Gritsev V and Demler E, \textit{Phys. Rev. Lett.} \textbf{104} (2010) 255302.

\bibitem{haller} Haller E, Gustavsson M, Mark M J, Danzl J G, Hart H, Pupillo G and N\"agerl H-C, \textit{Science} \textbf{325} (2009) 1224.

\bibitem{liao} Liao Y, Rittner C, Paprotta T, Li W, Partridge G B, Hulet R G, Baur S K and Mueller E J, \textit{Nature} \textbf{467} (2010) 567.

\bibitem{coldea} Coldea R, Tennant D A, Wheeler E M, Wawrzinska E, Prabhakaran D, Telling M, Habicht K, 
Smeibidil P and Kiefer K, \textit{Science} \textbf{327} (2010) 177.

\bibitem{nmr1} Fel'dman E B, Pyrkov A N, Zenchuk A I, \textit{Philosophical Transactions of The Royal Society A} \textbf{370} (2012) 4690.

\bibitem{hertier} H\'eritier M, \textit{Nature} \textbf{414} (2001) 31.

\bibitem{batchelor1} Batchelor M T, \textit{Physics Today} \textbf{60} (2007) 36.



\bibitem{batchelor3} Guan X-W , Batchelor M T and Lee C, \textit{Rev. Mod. Phys.} \textbf{85} (2013) 1633.

\bibitem{fst} Faddeev L D, Sklyanin E K and Takhtajan L A, \textit{Theor. Math. Phys.} \textbf{40} (1979) 194.

\bibitem{ks} Kulish P P and Sklyanin E K, \textit{Integrable Quantum Field Theories: 
Proceedings of the Symposium Held at Tvrminne, Finland - Lecture Notes in Physics} Editor: J. Hietarinta and C. Montonen, \textbf{151}, Springer-Verlag, Berlin, (1982) 61.

\bibitem{takhtajan} Takhtajan L A, \textit{Quantum Groups: 
Proceedings of the 8th International Workshop on Mathematical Physics Held at the Arnold Sommerfeld Institute, Clausthal, FRG - 
 Lecture Notes in Physics}, Editor: H. -D. Doebner and J. -D. Hennig, \textbf{370}, Springer-Verlag, Berlin, (1990) 3.


\bibitem{korepin} Korepin V E, Bogoliubov N M  and Izergin A G, \textit{Quantum inverse scattering method and correlation functions}, Cambridge University Press, Cambridge, (1993).


\bibitem{faddeev} Faddeev L D, \textit{Int. J. Mod. Phys. A } \textbf{10} (1995) 1845.

\bibitem{faddeev2} Faddeev L D, \textit{40 Years in Mathematical Physics - World Scientific Series in 20th Century Mathematics}, \textbf{2}, World Scientific Publishing Co. Pte. Ltd., Singapore, (1995).


\bibitem{korepin1} Izergin A G  and Korepin V E, \textit{Lett. Math. Phys.} \textbf{6} (1982) 283.

\bibitem{yang} Yang C N, \textit{Phys. Rev. Lett.} \textbf{19} (1967) 1312.

\bibitem{korepin2} Izergin A G  and Korepin V E, \textit{Nuc. Phys. B} \textbf{205} (1982) 401.

\bibitem{ek} Essler F H L  and Korepin V E, \textit{Exactly solvable models of strongly correlated electrons}, World Scientific, Singapore, (1994). 

\bibitem{ek2} Essler F H L, Frahm H, G\"ohmann F, Kl\"umper A and Korepin V E, \textit{The one-dimensional Hubbard Model}, Cambridge University Press, Cambridge, (2005). 

\bibitem{blz} Bazhanov V, Lukyanov S  and Zamolodchikov A B, \textit{Commun. Math. Phys.} \textbf{177} (1996) 381.

\bibitem{lipatov} Lipatov L, \textit{JETP Lett.} \textbf{59} (1994) 596.

\bibitem{korch} Faddeev L and Korchemsky G, \textit{Phys. Lett. B} \textbf{342} (1995) 311.

\bibitem{belitsky} Belitsky A V, Braun V M, Gorsky A S and Korchemsky G P, \textit{Int. J. Mod. Phys.} \textbf{A 19} (2004)  4715.

\bibitem{Gromov1} Escobedo J, Gromov N, Sever A and Vieira P, \textit{J. High Energy Phys.} \textbf{09} (2011) 028.

\bibitem{Gromov2} Gromov N, Vieira P, \textit{Phys. Rev. Lett.} \textbf{111} (2013) 211601.

\bibitem{CAhn} Beisert N, Ahn C, \textit{et al.}, \textit{Lett. Math. Phys.} \textbf{99} (2012) 3.

\bibitem{dorey}  Dorey N, \textit{J. Phys. A: Math. Theor.} \textbf{42} (2009) 254001.

\bibitem{jimbo85} Jimbo M, \textit{Lett. Math. Phys.} \textbf{10} (1985) 63. 

\bibitem{jimbo86} Jimbo M, \textit{Field Theory, Quantum Gravity and Strings: 
Proceedings of a Seminar Series Held at DAPHE, Observatoire de Meudon, and LPTHE, Universit\'e Pierre et Marie Curie, Paris - Lecture Notes in Physics}, Editor: H. J. de Vega and N. S\'anchez, \textbf{246}, Springer-Verlag, Berlin, (1986) 335.

\bibitem{drinfeld} Drinfeld V G, \textit{Quantum groups: Proc. Int. Congress of Mathematicians}, Editor: A. M. Gleason, Providence, RI: American Mathematical Society, (1986) 798.

\bibitem{frt} Reshetikhin N Yu, Takhtajan L A and Faddeev L D, \textit{Leningrad Math. J.} \textbf{1} (1990) 193.








\bibitem{Kuznet} Kuznetsov V B and Tsiganov A V, \textit{J. Phys. A: Math. Gen.} \textbf{22} (1989) L73.

\bibitem{CA-Gil} Santos Filho G. N., \textit{Physical and Mathematical Aspects of Symmetries}, Editor S. Duarte et al. (eds.), Springer International Publishing AG, (2017) 299.


\bibitem{roth} R. Roth and K. Burnett, \textit{Phys. Rev. A} \textbf{68} (2003)  023604-1.

\bibitem{jardel1} J. C. C. Cestari, A. Foerster and M. A. Gusmão, \textit{Phys. Rev. A} \textbf{82} (2010)  063634.

\bibitem{jardel2} J. C. C. Cestari,  A. Foerster, M. A. Gusmão and M. Continentino, \textit{Phys. Rev. A} \textbf{84} (2011)  055601.






\end{thebibliography}
\end{document}